\documentclass[journal]{IEEEtran}

\ifCLASSINFOpdf
\else
\fi
\usepackage{cite}
\usepackage{amssymb,amsfonts}
\usepackage{amsmath}
\usepackage{graphicx}
\usepackage{epstopdf}
\usepackage{color}
\usepackage{cases}
\usepackage{subfigure}
\usepackage{bm}

\usepackage{textcomp}
\usepackage{algorithm}
\usepackage{algorithmicx}

\newtheorem{assumption}{Assumption}
\newtheorem{lemma}{Lemma}
\newtheorem{theorem}{Theorem}

\newtheorem{remark}{Remark}

\newtheorem{corollary}{Corollary}

\begin{document}
\title{Linear Last-Iterate Convergence for Continuous Games with Coupled Inequality Constraints}
\author{
Min Meng and Xiuxian Li
\thanks{This work was partially supported by Shanghai Pujiang Program under Grant 21PJ1413100, the National Natural Science Foundation of China under Grant 62003243 and 62103305, Shanghai Municipal Science and Technology Major Project under Grant 2021SHZDZX0100, Shanghai Municipal Commission of Science and Technology under Grant 19511132101, Young Elite Scientist Sponsorship Program by cast of China Association for Science and Technology under Grant YESS20200136, and the Basic Science Centre Program by the National Natural Science Foundation of China under Grant 62088101.}
\thanks{M. Meng and X. Li are with the Department of Control Science and Engineering, College of Electronics and Information Engineering, and Shanghai Research Institute for Intelligent Autonomous Systems, Tongji University, Shanghai, China (Email: mengmin@tongji.edu.cn; xxli@ieee.org).
}
}

\maketitle

\begin{abstract}
In this paper, the generalized Nash equilibrium (GNE) seeking problem for continuous games with coupled affine inequality constraints is investigated in a partial-decision information scenario, where each player can only access its neighbors' information through local communication although its cost function possibly depends on all other players' strategies. To this end, a novel decentralized primal-dual algorithm based on consensus and dual diffusion methods is devised for seeking the variational GNE of the studied games. This paper also provides theoretical analysis to show that the designed algorithm converges linearly for the last-iterate, which, to our best knowledge, is the first to propose a linearly convergent GNE seeking algorithm under coupled affine inequality constraints. Finally, a numerical example is presented to demonstrate the effectiveness of the obtained theoretical results.
\end{abstract}


\IEEEpeerreviewmaketitle

\section{Introduction}
Game theory, which is the study of mathematical models of strategic interactions among rational agents, has been commonly applied in artificial intelligence (AI) \cite{zhao2022alphaholdem}, including multi-agent AI systems \cite{kraus2020ai}, imitation and reinforcement learning \cite{le2018hierarchical}, and adversarial training in generative adversarial networks \cite{wang2018kdgan}. A principal concept in noncooperative games is Nash equilibrium (NE), representing a stable and desirable state from which no one has incentive to deviate. The types of convergence involved in existing NE seeking  algorithms can be classified into the empirical distribution (i.e., time-average) of no-regret play and day-to-day play (i.e., last-iterate convergence) \cite{li2022survey}. In general, a no-regret learning algorithm can achieve sublinear regret bounds and generally ensure the convergence in the sense of empirical distribution, while may often fail to guarantee the last-iterate convergence, except for quite a few scenarios, such as two-player zero-sum games  \cite{hsieh2021adaptive}. As such, this paper focuses on the last-iterate convergence because of its significance in many practical applications, such as generative adversarial networks \cite{wang2018kdgan}.

On the other hand, decentralized Nash equilibrium (NE) and generalized NE (GNE) seeking problems have received considerable attention owing to 
the advantages of decentralized algorithms such as in scalability, reliability, robustness and efficiency.
Roughly speaking, the existing decentralized algorithms for seeking NEs of noncooperative games can be divided into three categories according to the types of constraints on the players' strategy sets. Specifically, the first class is for unconstrained games \cite{ye2017distributed,tatarenko2019geometric,zimmermann2021gradient}, i.e., each player can take actions from the whole set of real numbers or vectors, where a prerequisite on cost functions is required to ensure the existence of NEs. The second category is for games with local and uncoupled strategy set constraints  \cite{koshal2012gossip,salehisadaghiani2016distributed,zhu2021distributed,gadjov2019passivity,koshal2016distributed,salehisadaghiani2019distributed}, while the third type is for games with coupled constraints including affine/nonlinear equality/inequality constraints \cite{pavel2020distributed,zou2021continuous,meng2021decentralized,meng2022decentralized}. Most of these existing works focus on the design and the asymptotic convergence analysis of decentralized NE seeking algorithms, yet the convergence rate of the proposed algorithms is less discussed.

In fact, the convergence rate of an algorithm is extremely important in providing useful insights into how quickly the generated sequence approaches the target point or optimal solution. If the convergence rate is faster, then fewer iterations can possibly yield a good approximation of the optimal solution within a required error, while it cannot see any conclusive information from any finite part of an asymptotically convergent sequence. Therefore, proposing a decentralized NE seeking algorithm with a faster convergence rate is necessary.

{\bf Our Contribution.}
In this paper, we consider the GNE seeking problem for continuous games with coupled affine inequality constraints, where individual players aim to privately minimize their own cost functions by selecting a strategy profile satisfying the coupled inequality. It is noted that the cost function of each player may depend on all other players' strategies, and it is possibly impractical for all information sharing especially in large-scale networks. Hence, a partial-decision information scenario, i.e., each player can only access its own strategy and cost function, as well as neighbors' strategies through communicating via a connected graph, is considered as done in some existing works \cite{zhu2021distributed,salehisadaghiani2019distributed,zou2021continuous,meng2021decentralized,meng2022decentralized,tatarenko2018geometric,bianchi2020fully,bianchi2022fast,zhu2021generalized,pavel2020distributed}. In such setting, a novel decentralized primal-dual algorithm, where each player is equipped with additional variables to estimate all the other players' actions and the global dual variable, is proposed to learn the unique variational GNE of the considered game and is also strictly proved to be linearly convergent for the last-iterate.
Finally, a numerical example on a Nash-Cournot game is presented to illustrate the effectiveness of theoretical results.
The main contributions of this paper are summarized as follows.
\begin{itemize}
\item[1)] To the best of our knowledge, this paper is the first to propose a linearly last-iterate convergent decentralized GNE seeking algorithm for games with coupled affine inequality constraints. It can also be seen from simulations that our designed algorithm outperforms the forward-backward (FB) algorithm proposed in \cite{pavel2020distributed} in terms of the convergence rate. Furthermore, the upper bounds for required stepsizes are explicitly provided, which depend on the number of players, the communication structure, and the affine inequality along with the Lipschitz and monotonicity constants of the pseudo-gradient function.
\item[2)] The linearly convergent algorithm is creatively designed by modifying the typical primal-dual algorithm and drawing into a positive semi-definite matrix related to the communication topology, and simultaneously following the consensus and dual diffusion methods. A critical and challenging step to derive the linear convergence rate of the presented algorithm is elaborately constructing a weighted error norm sum.
\end{itemize}

{\bf Related Work.}
To date, it has been proven that decentralized gradient-based algorithms to seek NEs for continuous games without constraints converge linearly \cite{tatarenko2019geometric,zimmermann2021gradient}. For continuous games with local strategy set constraints, an inexact-ADMM algorithm was proposed in \cite{salehisadaghiani2019distributed} for finding NEs, which was proved to converge relatively fast with a sublinear convergence rate $O(\frac{1}{k})$, where $k$ is the iteration time. Then, decentralized algorithms based on gradient-play method were devised for games with local set constraints in \cite{tatarenko2018geometric,bianchi2020fully} and were shown to possess a linear convergence rate. Nevertheless, there are few works on linearly convergent GNE seeking algorithms for games with coupled constraints. A recent study \cite{bianchi2022fast} proposed a proximal-point algorithm by designing a novel preconditioning matrix for learning GNE of games with strategy set constraints and affine inequality constraints, which can improve the convergence speed compared with gradient-play methods, yet the linear convergence is not established. In \cite{zhu2021generalized}, a continuous-time GNE learning algorithm in continuous-time for continuous games with strategy set constraints and affine equality constraints was studied and theoretically shown to have an exponential convergence rate when no local strategy set constraints are involved.
Therefore, the linear convergence of existing NE (GNE) seeking algorithms is not analyzed except for the ones under only local strategy set constraints, and our paper is the first to design a linearly convergent GNE seeking algorithm under coupled constraints.

The rest of this paper is organized as follows. Section \ref{section2} introduces the problem formulation and Section \ref{section3} presents the main results on the design and convergence analysis of the proposed algorithm. In Section \ref{section4}, simulations are provided to demonstrate the effectiveness of the devised algorithm. Finally, a brief conclusion is drawn in Section \ref{section5}.

{\bf Notations.} Denote by $\mathbb{R}$, $\mathbb{R}^n$ and $\mathbb{R}^{m\times n}$ the sets of real numbers, real $n$-dimensional column vectors and real $n\times n$-matrices, respectively. Let $\mathbb{R}^n_+$ be the set of nonnegative real vectors. The symbol $[m]$ for an integer $m>0$ represents the set $\{1,2,\ldots,m\}$. ${\bf 1}_n\in\mathbb{R}^n$ (resp. ${\bf 0}_n\in\mathbb{R}^n$) is a vector with all elements being 1 (resp. 0). For a vector or matrix $A$, its transpose is denoted as $A^{\top}$. $col(x_1,\ldots,x_n):=(x_1^{\top},\ldots,x_n^{\top})^{\top}$. The Kronecker product of matrices $A$ and $B$ is denoted as $A\otimes B$. $P_{\Omega}[x]:=\arg\min_{y\in\Omega}\|y-x\|$ denotes the projection of the vector $x\in\mathbb{R}^n$ onto the closed convex set $\Omega$ and $N_{\Omega}(x):=\{v|v^{\top}(y-x)\leq 0,~\forall y\in\Omega\}$ is the normal cone to $\Omega$ at $x\in\Omega$.
${\rm diag}\{a_1,a_2,\ldots,a_n\}$ represents a diagonal matrix with $a_i$, $i\in[n]$, on its diagonal. Denote by $\lambda_{\min}(A)$, $\lambda_{\max}(A)$ and $\rho(A)$ the smallest eigenvalue, largest eigenvalue and spectral radius of a square matrix $A$, respectively. $\underline{\sigma}(A)$ is the smallest nonzero singular value of $A$. For a positive semi-definite matrix $M\in\mathbb{R}^{n\times n}$, denote $\|x\|_{M}:=\sqrt{x^{\top}Mx}$ with $x\in\mathbb{R}^n$.

\section{Problem Formulation}\label{section2}
Consider a normal-form continuous game with $N$ players, where each player takes its strategy (decision, action) $x_i\in\mathbb{R}^{n_i}$. Denote by $x=col(x_1,\ldots,x_N)$ and $x_{-i}:=col(x_1,\ldots,x_{i-1},x_{i+1},\ldots,x_N)$ the joint action of all the players and the joint action of all the players except $i$, respectively. The private cost function of player $i$ is $f_i(x_i,x_{-i})$, depending on both local variable $x_i$ and other players' decisions $x_{-i}$. The objective of player $i$ is to selfishly minimize its own cost function $f_i(x_i,x_{-i})$ subject to a coupled constraint $\Omega:=\{x\in\mathbb{R}^n|Ax\leq b\}$, where $A:=[A_1,A_2,\ldots,A_N]$, $b:=\sum_{i=1}^Nb_i$, $n:=\sum_{i=1}^Nn_i$, $A_i\in\mathbb{R}^{m\times n_i}$ and $b_i\in\mathbb{R}^{m}$. Here, $A_i$ and $b_i$ can only be privately accessible to player $i$. In this setting, a strategy profile $x^*=(x_i^*,x_{-i}^*)$ is called a GNE if for each $i\in[N]$, the following inequality holds:
\begin{align}\label{equ1}
f_i(x_i^*,x_{-i}^*)\leq f_i(x_i,x_{-i}^*), ~\forall x_i\in X_i(x_{-i}^*),
\end{align}
where $X_i(x_{-i}^*):=\{x_i\in\mathbb{R}^{n_i}|A_ix_i\leq b-\sum_{j\neq i}A_jx_j^*\}$.

Some standard assumptions on this game are listed as follows.
\begin{assumption}\label{assumption1}
For each $i\in[N]$, the function $f_i(x_i,x_{-i})$ is continuously
differentiable and convex with respect to $x_i$ given any $x_{-i}$. For each $i\in[N]$, $\nabla_if_i(x_i,x_{-i}):=\frac{\partial f_i(x_i,x_{-i})}{\partial x_i}$ is $L$-Lipschitz continuous for some $L>0$, i.e.,
\begin{align*}
    \|\nabla_if_i(x)-\nabla_if_i(y)\|\leq L\|x-y\|,~\forall x,y\in\mathbb{R}^{n}.
\end{align*}
Moreover, the constraint set $\Omega$ is nonempty and satisfies Slater's constraint qualification.
\end{assumption}

Define the pseudo-gradient $F:\mathbb{R}^n\to\mathbb{R}^n$ of game (\ref{equ1}) as
\begin{align}\label{equ2}
F(x):=col(\nabla_1f_1(x_1,x_{-1}),\ldots,\nabla_Nf_N(x_N,x_{-N})).
\end{align}

\begin{assumption}\label{assumption2}
The pseudo-gradient $F$ defined in (\ref{equ2}) is $\mu$-strongly monotone for some $\mu>0$, that is, there holds
\begin{align*}
(F(x)-F(y))^{\top}(x-y)&\geq\mu\|x-y\|^2,~\forall x,y\in\mathbb{R}^{n}.
\end{align*}
\end{assumption}

The strong monotonicity of $F$ is critical to ensure linear convergence of decentralized algorithms with fixed step-sizes.

From the definition of GNE in (\ref{equ1}), the strategy profile $x^*=(x_i^*,x_{-i}^*)$ is a GNE, if and only if $x_i^*$ is an optimal solution to the following optimization problem
\begin{align*}
&\min_{x_i\in\mathbb{R}^{n_i}} ~~f_i(x_i,x_{-i}^*)\\
&~~~{\rm s.t.}~~~~   A_ix_i\leq b-\sum_{j\neq i}A_jx_j^*.
\end{align*}
For this optimization problem, the Lagrangian function for each player $i\in[N]$ is given as
\begin{align}\label{equ3}
\mathcal{L}_i(x_i,\lambda_i;x_{-i}):=f_i(x_i,x_{-i})+\lambda_i^{\top}(Ax-b),
\end{align}
where $\lambda_i\in\mathbb{R}^m_+$ is the dual variable. If the profile $x^*=(x_i^*,x_{-i}^*)$ is a GNE, then by Karush-Kuhn-Tucker (KKT) conditions of this optimization problem, there exist dual variables $\lambda_i^*\in\mathbb{R}^m_+$, $i\in[N]$, such that
\begin{align}
{\bf 0}_{n_i}&=\nabla_if_i(x_i^*,x_{-i}^*)+A_i^{\top}\lambda_i^*,\label{equ4} \\
{\bf 0}_m&\in-(Ax^*-b)+N_{\mathbb{R}^m_+}(\lambda_i^*),~i\in[N].\label{equ5}
\end{align}

If $\lambda_1^*=\lambda_2^*=\cdots=\lambda_N^*=\lambda^*$, then the GNE with dual variable $\lambda^*$ is called a variational GNE,  as discussed in many existing references \cite{pavel2020distributed,zou2021continuous,meng2021decentralized,meng2022decentralized}, which is economically justifiable. Here, it is assumed that a variational GNE of the considered game exists as we focus on the design and linear convergence analysis of the
GNE seeking algorithm rather than the existence of GNEs. From the variational inequality perspective, a variational GNE $x^*$ is a solution to the following variational inequality:
\begin{align*}
F^{\top}(x^*)(x-x^*)\geq0,~\forall x\in\Omega.
\end{align*}
Then, under Assumption \ref{assumption2}, game (\ref{equ1}) has a unique variational GNE \cite{facchinei2003finite}.

On the other hand, owing to the partial-decision information setting, the communication pattern among all players is depicted by an undirected graph $\mathcal{G}=(\mathcal{V},\mathcal{E},W)$, where $\mathcal{V}=[N]$ is the node set corresponding to all players, $\mathcal{E}$ is the edge set, and $W=(w_{ij})\in\mathbb{R}^{N\times N}$ is the adjacency matrix. $w_{ij}>0$ if $(i,j)\in\mathcal{E}$ meaning that players $i$ and $j$ can communicate directly with each other, and otherwise $w_{ij}=0$. It is supposed that $w_{ii}>0$ for all $i\in[N]$ in this paper. A standard assumption is made on graph $\mathcal{G}$.
\begin{assumption}\label{assumption3}
$\mathcal{G}$ is connected. The adjacency matrix $W$ is symmetric and doubly stochastic, that is, $W^{\top}=W$, $W{\bf1}_N={\bf1}_N$ and ${\bf1}_N^{\top}W={\bf1}_N^{\top}$.
\end{assumption}

Assumption \ref{assumption3} is a mild condition in decentralized algorithms, and from Assumption \ref{assumption3}, one has
\begin{align}\label{equ6}
    \sigma:=\|W-\mathbf{1}_N\mathbf{1}_N^{\top}/N\|\in[0,1).
\end{align}

\begin{assumption}\label{assumption4}
For each $i\in[N]$, $A_i$ has full row rank.
\end{assumption}

Under Assumption \ref{assumption4}, it can be ensured that $\lambda_{\min}(A_iA_i^{\top})>0$. Assumption \ref{assumption4} is essential in proving the linear convergence of the designed algorithm, which is often made for linear convergent algorithms for decentralized optimization with affine constraints (see \cite{alghunaim2021dual} and references therein).

In summary, in this setting, the aim of this paper is to design a decentralized algorithm to find the unique variational GNE of game (1) based on local information and also establish the linear convergence rate of the designed algorithm.

\section{The Developed Algorithm and Convergence Analysis}  \label{section3}
In this section, a novel decentralized primal-dual algorithm is first proposed to learn the unique variational GNE for coupled constrained games in a partial-decision information scenario and then the linear convergence result is provided.

Note that each player cannot access the information of all other players in the partial-decision information setting. Then, each player is assigned two additional variables to estimate the decisions of other players and the global dual variable, respectively, through local communication via the graph $\mathcal{G}$.
It can be obtained that the weighted matrix $W$ is primitive under Assumption \ref{assumption3}. Then, $W$ has a simple eigenvalue 1 and other eigenvalues being in $(-1,1)$. Therefore, $I_N-W$ is a positive semi-definite matrix, and $(I_{Nm}-W\otimes I_m){\bm\lambda}=0$ for ${\bm\lambda}\in\mathbb{R}^{Nm}$ if and only if ${\bm\lambda}={\bf 1}_N\otimes\lambda$ for some $\lambda\in\mathbb{R}^{m}$. Define a symmetric matrix $B\in\mathbb{R}^{N\times N}$ such that
\begin{align*}
B^2=\frac{1}{2}(I_{N}-W),
\end{align*}
then $\lambda_{\min}({B}^2)=0$, $\lambda_{\max}({B}^2)=\lambda_{\max}^2({B})<1$, and it holds that $({B}\otimes I_m){\bm\lambda}=0$ for ${\bm\lambda}\in\mathbb{R}^{Nm}$ is equivalent to ${\bm\lambda}={\bf1}_N\otimes\lambda$ for some $\lambda\in\mathbb{R}^m$.

Denote $\Pi:={\rm diag}\{A_1,\ldots,A_N\}$, ${\bf b}:=col(b_1,\ldots,b_N)$ and $\mathcal{B}:=B\otimes I_m$. At iteration $k$, player $i$ is endowed with variables $x_{i,k}$, $x_{ji,k}$ and $\lambda_{i,k}$ to represent its decision, the estimate of player $j$' decision and the estimate of the global dual variable, respectively. $x_{ii,k}=x_{i,k}$.  Then, a decentralized primal-dual algorithm is proposed as follows:
\begin{subequations}\label{equ7}
\begin{align}
x_{i,k+1}&=\sum_{j=1}^Nw_{ij}x_{ij,k}-\alpha\nabla_if_i(x_{i,k},{\bf x}_{-i,k})\nonumber\\
&~~~-\alpha A_i^{\top}\lambda_{i,k},\label{equ7a}\\
{\bf x}_{-i,k+1}&=\sum_{j=1}^Nw_{ij}{\bf x}_{-i,k}^j,\label{equ7b}\\
\mathbf{v}_{k+1}&={\bm\lambda}_k-\mathcal{B}^2{\bm\lambda}_k+\beta(\Pi x_{k+1}-{\bf b})+\mathcal{B}\mathbf{y}_k,\label{equ7c}\\
\mathbf{y}_{k+1}&=\mathbf{y}_k-\gamma\mathcal{B}\mathbf{v}_{k+1},\label{equ7d}\\
{\bm\lambda}_{k+1}&=P_{\mathbb{R}^{Nm}_+}[\mathbf{v}_{k+1}],\label{equ7e}
\end{align}
\end{subequations}
where ${\bf x}_{-i,k}:=col(x_{1i,k},\ldots,x_{(i-1)i,k},x_{(i+1)i,k},\ldots,x_{Ni,k})$, ${\bf x}_{-i,k}^j:=col(x_{1j,k},\ldots,x_{(i-1)j,k},x_{(i+1)j,k},\ldots,x_{Nj,k})$, $x_k=col(x_{1,k},\ldots,x_{N,k})$, ${\bm\lambda}_k:=col(\lambda_{1,k},\ldots,\lambda_{N,k})$, $\mathbf{v}_k,\mathbf{y}_k\in\mathbb{R}^{Nm}$ are auxiliary variables, and $\alpha,\beta,\gamma>0$ are stepsizes to be determined. Moreover, one can initialize $x_{i,0}\in\mathbb{R}^{n_i}$, ${\bf x}_{-i,0}\in\mathbb{R}^{n-n_i}$, ${\bm\lambda_0}\in\mathbb{R}^{Nm}$  arbitrarily, and $\mathbf{y}_0={\bf0}_{Nm}$. Iteration (\ref{equ7}) cannot be implemented in a fully decentralized manner since the matrix $B$ is involved. In what follows, let us equivalently transfer iteration (\ref{equ7}) into a fully decentralized algorithm. By (\ref{equ7c}), one has
\begin{align}
\mathbf{v}_{k+1}-\mathbf{v}_k&=(I_{Nm}-\mathcal{B}^2)({\bm\lambda}_k-{\bm\lambda}_{k-1})+\beta\Pi(x_{k+1}-x_{k})\nonumber\\
&~~~~+\mathcal{B}(\mathbf{y}_k-\mathbf{y}_{k-1}).\label{equ8}
\end{align}
Substituting (\ref{equ7d}) into (\ref{equ8}) yields that for $k\geq0$,
\begin{align}
\mathbf{v}_{k+1}&=(I_{Nm}-\gamma\mathcal{B}^2)\mathbf{v}_k+(I_{Nm}-\mathcal{B}^2)({\bm\lambda}_k-{\bm\lambda}_{k-1})\nonumber\\
&~~~~+\beta\Pi(x_{k+1}-x_{k}),\label{equ9}
\end{align}
where $\mathbf{v}_0$, ${\bm\lambda}_{-1}$ and $x_{-1}$ are set to be $\mathbf{v}_0={\bf0}_{Nm}$, ${\bm\lambda}_{-1}={\bf0}_{Nm}$ and $\Pi x_{-1}={\bf b}$, respectively. Therefore, iteration (\ref{equ7}) can be rewritten as a fully decentralized algorithm (cf. Algorithm \ref{alg1}).

\begin{algorithm}[!htbp]\caption{\bf Decentralized Primal-Dual Algorithm}\label{alg1}
Each player $i$ maintains vector variables $x_{i,k}\in\mathbb{R}^{n_i}$, $x_{ji,k}\in\mathbb{R}^{n_j}$, $v_{i,k}\in\mathbb{R}^m$ and $\lambda_{i,k}\in\mathbb{R}^{m}$ at iteration $k$. Let $C=(c_{ij})_{N\times N}:=B^2=\frac{1}{2}(I_N-W)$.\\
{\bf Initialization:} For any $i\in[N]$, initialize $x_{i,0}$, $x_{ji,0}$ and $\lambda_{i,0}$ arbitrarily, and set $v_{i,0}={\bf 0}_m$, $\lambda_{i,-1}={\bf0}_m$ and $A_ix_{i,-1}=b_i$.\\
{\bf Iteration:} For every player $i$, repeat for $k\geq0$:
\begin{subequations}\label{equ10}
\begin{align}
x_{i,k+1}&=\sum_{j=1}^Nw_{ij}x_{ij,k}-\alpha\nabla_if_i(x_{i,k},{\bf x}_{-i,k})-\alpha A_i^{\top}\lambda_{i,k},\label{equ10a}\\
{\bf x}_{-i,k+1}&=\sum_{j=1}^Nw_{ij}{\bf x}_{-i,k}^j,\label{equ10b}\\
v_{i,k+1}&=v_{i,k}-\gamma\sum_{j=1}^Nc_{ij}v_{j,k}-\sum_{j=1}^Nc_{ij}(\lambda_{j,k}-\lambda_{j,k-1})\nonumber\\
&~~~+\lambda_{i,k}-\lambda_{i,k-1}+\beta A_i(x_{i,k+1}-x_{i,k}),\label{}\\
{\lambda}_{i,k+1}&=P_{\mathbb{R}^{m}_+}[v_{i,k+1}].
\end{align}
\end{subequations}
\end{algorithm}

\begin{remark}
Algorithm (\ref{equ7}) is inspired by \cite{alghunaim2019linearly,alghunaim2021dual}, where the aim is to minimize a global cost function. However, the problem studied here is different, i.e., continuous games, which, together with partial-decision information setting, leads to that the theoretical analysis in this paper is significantly distinctive from that in \cite{alghunaim2019linearly,alghunaim2021dual}.
\end{remark}

\begin{remark}
Algorithm \ref{alg1}, which is equivalent to iteration (\ref{equ7}), is not a typical primal-dual method since the auxiliary variable $\mathbf{v}_k$, instead of the dual variable ${\bm\lambda}_k$, is used in (\ref{equ7d}). This, together with the term $-\mathcal{B}^2\bm{\lambda}_k$ used in (\ref{equ7c}), plays a critical role in establishing linear convergence when dealing with coupled inequality constraints. In fact, $\mathbf{v}_k$ used in (\ref{equ7d}) can help derive an equality on $\mathbf{v}_{k+1}$ and $\bm{\lambda}_k$, and having the term $-\mathcal{B}^2\bm{\lambda}_k$ in (\ref{equ7c}) is equivalent to adding $-1/2\bm{\lambda}^{\top}\mathcal{B}^2\bm{\lambda}$ to the Lagrangian function (\ref{equ3}) since $\mathcal{B}\bm{\lambda}^*=\mathbf{0}_{Nm}$. Moreover, note that the projection step in (\ref{equ7e}) is to tackle the inequality constraints, then Algorithm \ref{alg1} can reduce to the case of affine equality constraints by letting $\bm{\lambda}_k=\bf{v}_k$ and the results obtained in this paper are applicable to solving the distributed GNE seeking problem with coupled affine equality constraints.
\end{remark}

Next, we rewrite (\ref{equ10a}) and (\ref{equ10b}) into a compact form by introducing two matrices $\mathcal{R}_i$ and $\mathcal{S}_i$ for player $i$ to manipulate its decision variable $x_{i,k}$ and the estimate ${\bf x}_{-i,k}$ of other players' decisions. Let
\begin{align}
\mathcal{R}_i&:=\left[
\begin{array}{ccc}
\mathbf{0}_{n_i\times n_{<i}}&I_{n_i}&\mathbf{0}_{n_i\times n_{>i}}
\end{array}
\right],\label{equ11}\\
\mathcal{S}_i&:=\left[
\begin{array}{ccc}
I_{n_{<i}}&{\bf0}_{n_{<i}\times n_i}&{\bf0}_{n_{<i}\times n_{>i}}\\
{\bf0}_{n_{>i}\times n_{<i}}&{\bf0}_{n_{>i}\times n_i}&I_{n_{>i}}
\end{array}
\right],
\end{align}
where $n_{<i}:=\sum_{j=1}^{i-1}n_j$ and $n_{>i}:=\sum_{j=i+1}^{N}n_j$. Then it can be easily verified that $\mathcal{R}:={\rm diag}\{\mathcal{R}_1,\ldots,\mathcal{R}_N\}$ and $\mathcal{S}:={\rm diag}\{\mathcal{S}_1,\ldots,\mathcal{S}_N\}$ satisfy
\begin{align*}
&\mathcal{R}\mathcal{R}^{\top}=I_n,~\mathcal{S}\mathcal{S}^{\top}=I_{Nn-n},\\
&\mathcal{R}^{\top}\mathcal{R}+\mathcal{S}^{\top}\mathcal{S}=I_{Nn},\\
&\mathcal{R}\mathcal{S}^{\top}={\bf0},~\mathcal{S}\mathcal{R}^{\top}={\bf0}.
\end{align*}
Denote
\begin{align*}
{\bf x}_{i,k}&:=col(x_{1i,k},\ldots,x_{Ni,k}),\\
{\bf x}_k&:=col({\bf x}_{1,k},\ldots,{\bf x}_{N,k}).
\end{align*}
One has
\begin{align}
\mathcal{R}_i{\bf x}_{i,k}&=x_{i,k},\\
\mathcal{S}_i{\bf x}_{i,k}&={\bf x}_{-i,k}.
\end{align}
Hence, $x_k=\mathcal{R}{\bf x}_k$ and $col({\bf x}_{-1,k},\ldots,{\bf x}_{-N,k})=\mathcal{S}{\bf x}_k$. With these notations, (\ref{equ10a}) and (\ref{equ10b}) can be rewritten as
\begin{align}
x_{k+1}&=\mathcal{R}(W\otimes I_n){\bf x}_k-\alpha{\bf F}({\bf x}_k)-\alpha\Pi^{\top}{\bm\lambda}_{k},\\
\mathcal{S}{\bf x}_{k+1}&=\mathcal{S}(W\otimes I_n){\bf x}_{k},
\end{align}
where
\begin{align}\label{equ17}
&{\bf F}({\bf x}_k)\nonumber\\
&:=col(\nabla_1f_1(x_{1,k},{\bf x}_{-1,k}),\ldots,\nabla_Nf_N(x_{N,k},{\bf x}_{-N,k})).
\end{align}
In view of ${\bf x}_k=\mathcal{R}^{\top}x_k+\mathcal{S}^{\top}\mathcal{S}{\bf x}_k$, the iteration of ${\bf x}_k$ can be derived as
\begin{align}
{\bf x}_{k+1}=(W\otimes I_n){\bf x}_k-\alpha\mathcal{R}^{\top}{\bf F}({\bf x}_k)-\alpha\mathcal{R}^{\top}\Pi^{\top}{\bm\lambda}_k.\label{equ18}
\end{align}

Before presenting the main result, some auxiliary lemmas are first provided.
\begin{lemma}\label{lemma1}
Iteration (\ref{equ10}) or equivalently (\ref{equ7}) has a fixed point $({\bf x}^*,\mathbf{v}^*,\mathbf{y}^*,{\bm\lambda}^*)$ satisfying
\begin{align}
    {\bf x}^*&=(W\otimes{I_n})\mathbf{x}^*-\alpha\mathcal{R}^{\top}\mathbf{F}(\mathbf{x}^*)-\alpha\mathcal{R}^{\top}\Pi^{\top}{\bm\lambda}^*,\label{equ19}\\
    \mathbf{v}^*&={\bm\lambda}^*+\beta(\Pi x^*-{\bf b})+\mathcal{B}\mathbf{y}^*,\label{equ20}\\
    \mathcal{B}\mathbf{v}^*&=\mathbf{0}_{Nm},\label{equ21}\\
    {\bm\lambda}^*&=P_{\mathbb{R}^{Nm}_+}[\mathbf{v}^*],\label{equ22}
\end{align}
and $\mathcal{B}^2{\bm\lambda}^*=\mathbf{0}_{Nm}$. For any fixed point $({\bf x}^*,\mathbf{v}^*,\mathbf{y}^*,{\bm\lambda}^*)$, it holds that $\mathbf{x}^*={\bf1}_N\otimes x^*$ for $x^*\in\mathbb{R}^{n}$ and ${\bm\lambda}^*=\mathbf{1}_{N}\otimes\lambda^*$ for $\lambda^*\in\mathbb{R}^m$ with $x^*$ being the variational GNE of game (\ref{equ1}) and $\lambda^*$ being the optimal global dual variable.
\end{lemma}

\emph{Proof:} See Appendix. \hfill$\blacksquare$

For the fixed point of iteration (\ref{equ7}), $(\mathbf{x}^*,\mathbf{v}^*,\mathbf{y}^*,{\bm\lambda}^*)$ with $\mathbf{x}^*=\mathbf{1}_N\otimes{x}^*$, $\mathbf{v}^*=\mathbf{1}_N\otimes{v}^*$, ${\bm\lambda}^*={\bf1}_N\otimes{\lambda}^*$ and $\mathbf{y}^*$ being in the range space of $\mathcal{B}$, define the error variables as follows:
\begin{align}
    &\tilde{\bf x}_k:=\mathbf{x}_k-\mathbf{x}^*,~~\tilde{\mathbf{v}}_k:=\mathbf{v}_k-\mathbf{v}^*,\\
    &\tilde{\mathbf{y}}_k:=\mathbf{y}_k-\mathbf{y}^*,~~\tilde{\bm\lambda}_k:={\bm\lambda}_k-{\bm\lambda}^*.
\end{align}
Based on (\ref{equ7}) and (\ref{equ18})--(\ref{equ22}), the error variables evolve as
\begin{align}
    \tilde{\bf x}_{k+1}&=(W\otimes I_n)\tilde{\bf x}_k-\alpha\mathcal{R}^{\top}({\bf F}({\bf x}_k)-{\bf F}({\bf x}^*))\nonumber\\
    &~~~-\alpha\mathcal{R}^{\top}\Pi^{\top}\tilde{\bm\lambda}_k,\label{equ29}\\
\tilde{\mathbf{v}}_{k+1}&=\tilde{\bm\lambda}_k-\mathcal{B}^2\tilde{\bm\lambda}_k+\beta\Pi(x_{k+1}-x^*)+\mathcal{B}\tilde{\mathbf{y}}_k,\label{equ30}\\
\tilde{\mathbf{y}}_{k+1}&=\tilde{\mathbf{y}}_k-\gamma\mathcal{B}\tilde{\mathbf{v}}_{k+1},\label{equ31}\\
\tilde{\bm\lambda}_{k+1}&=P_{\mathbb{R}^{Nm}_+}[\mathbf{v}_{k+1}]-P_{\mathbb{R}^{Nm}_+}[\mathbf{v}^*].\label{equ32}
\end{align}
Denote $\mathcal{W}:=W\otimes{I_n}$, $\mathcal{W}_{\infty}:={\bf1}_N{\bf1}_N^{\top}/N\otimes{I_n}$ and $\mathbf{x}_{\bot,k}:=(I_{Nn}-\mathcal{W}_{\infty})\mathbf{x}_k$, then $\mathbf{x}_k=\mathcal{W}_{\infty}\mathbf{x}_k+\mathbf{x}_{\bot,k}$, $(\mathbf{x}_{\bot,k})^{\top}\mathcal{W}_{\infty}\mathbf{x}_k=0$, and $\mathcal{W}\mathcal{W}_{\infty}=\mathcal{W}_{\infty}\mathcal{W}=\mathcal{W}_{\infty}\mathcal{W}_{\infty}=\mathcal{W}_{\infty}$.
Hence,
\begin{align}
    \|\mathcal{W}\mathbf{x}_k-\mathbf{x}^*\|^2
    &=\|\mathcal{W}\mathcal{W}_{\infty}\mathbf{x}_k-\mathbf{x}^*+\mathcal{W}\mathbf{x}_{\bot,k}\|^2\nonumber\\
    &=\|\mathcal{W}_{\infty}\mathbf{x}_k-\mathbf{x}^*\|^2+\|\mathcal{W}\mathbf{x}_{\bot,k}\|^2,\label{equ35}\\
    \|\mathcal{W}\mathbf{x}_{\bot,k}\|&=\|(\mathcal{W}-\mathcal{W}_{\infty})\mathbf{x}_{\bot,k}\|\nonumber\\
    &\leq\sigma\|\mathbf{x}_{\bot,k}\|,\label{equ36}
\end{align}
where the inequality is derived based on (\ref{equ6}).

\begin{lemma}\label{lemma2}
Under Assumptions \ref{assumption1}--\ref{assumption3}, the error variable $\tilde{\mathbf{x}}_k$ generated by Algorithm \ref{alg1} satisfies
\begin{align}
    \|\tilde{\mathbf{x}}_{k+1}\|^2&\leq\rho(M_{\alpha})\|\tilde{\mathbf{x}}_{k}\|^2-\frac{\mu}{N}\alpha\|\mathcal{W}_{\infty}\mathbf{x}_k-\mathbf{x}^*\|^2\nonumber\\
    &~~~-\sigma{L}\alpha\|\mathbf{x}_{\bot,k}\|^2-\alpha^2\|\mathcal{R}^{\top}\Pi^{\top}\tilde{\bm\lambda}_k\|^2\nonumber\\
    &~~~+2\alpha(x^*-x_{k+1})^{\top}\Pi^{\top}\tilde{\bm\lambda}_k,
\end{align}
where
\begin{align}
    M_{\alpha}:=\left[
    \begin{array}{ccc}
        1-\frac{\mu}{N}\alpha+L^2\alpha^2 & (\sigma+1)L\alpha \\
        (\sigma+1)L\alpha & \sigma^2+3\sigma L\alpha+L^2\alpha^2
    \end{array}
    \right].
\end{align}
\end{lemma}

\emph{Proof:} See Appendix.
\hfill$\blacksquare$

\begin{lemma}\label{lemma3}
The error variables $\tilde{\mathbf{v}}_k$, $\tilde{\mathbf{y}}_k$ and $\tilde{\bm\lambda}_k$ generated by Algorithm \ref{alg1} satisfy
\begin{align}
    &\|\tilde{\mathbf{v}}_{k+1}\|^2_{I_{Nm}-\gamma\mathcal{B}^2}+\gamma^{-1}\|\tilde{\mathbf{y}}_{k+1}\|^2\nonumber\\
    &=\|\tilde{\bm\lambda}_k-\mathcal{B}^2\tilde{\bm\lambda}_k+\beta\Pi(x_{k+1}-x^*)\|^2-\|\mathcal{B}\tilde{\mathbf{y}}_k\|^2\nonumber\\
    &~~~+\gamma^{-1}\|\tilde{\mathbf{y}}_k\|^2,\label{equ41}
\end{align}
where $\gamma>0$ is chosen as $\gamma<\lambda^{-2}_{\max}(\mathcal{B})$.
\end{lemma}

\emph{Proof:} See Appendix.
\hfill$\blacksquare$

Next, it is ready to present the main convergence result on Algorithm \ref{alg1}.
\begin{theorem}\label{theorem1}
Under Assumptions \ref{assumption1}--\ref{assumption4}, the sequences $\{\mathbf{x}_k\}$, $\{\mathbf{v}_k\}$, $\{\mathbf{y}_k\}$ and $\{{\bm\lambda}_k\}$ generated by Algorithm \ref{alg1} satisfy
\begin{align}
    E(k+1)&
    \leq\rho(M_{\alpha})\|\tilde{\mathbf{x}}_k\|^2-\min\{\mu/N,\sigma L\}\alpha\|\tilde{\mathbf{x}}_k\|^2\nonumber\\
    &~~~-\alpha^2\|\Pi^{\top}\tilde{\bm\lambda}_k\|^2+\alpha\beta^{-1}\|\tilde{\bm\lambda}_k\|^2+2\alpha\beta^{-1}\|\mathcal{B}^2\tilde{\bm\lambda}_k\|^2\nonumber\\
    &~~~-2\alpha\beta^{-1}\|\mathcal{B}\tilde{\bm\lambda}_k\|^2-\alpha\beta^{-1}\|\mathcal{B}\tilde{\mathbf{y}}_k\|^2\nonumber\\
    &~~~+\alpha\beta^{-1}\gamma^{-1}\|\tilde{\mathbf{y}}_k\|^2,\label{equ45}
\end{align}
where $E(k+1):=\|\tilde{\mathbf{x}}_{k+1}\|^2_{I_{Nn}-2\alpha\beta\mathcal{R}^{\top}\Pi^{\top}\Pi\mathcal{R}}+\alpha\beta^{-1}\|\tilde{\bm\lambda}_{k+1}\|^2_{I_{Nm}-\gamma\mathcal{B}^2}+\alpha\beta^{-1}\gamma^{-1}\|\tilde{\mathbf{y}}_{k+1}\|^2$.

In addition, if the stepsizes $\alpha,\beta,\gamma>0$ are chosen such that $\rho(M_{\alpha})<1$ and
\begin{align}
    \beta&<\min\Big\{\frac{\mu}{2N\lambda_{\max}(\Pi^{\top}\Pi)},\frac{\sigma L}{2\lambda_{\max}(\Pi^{\top}\Pi)},\frac{1}{\alpha\lambda_{\min}(\Pi\Pi^{\top})}\Big\},\nonumber\\
    \gamma&<\min\Big\{\frac{2-2\lambda_{\max}^2(\mathcal{B})}{1-\alpha\beta\lambda_{\min}(\Pi\Pi^{\top})},\frac{1}{\lambda_{\max}^2(\mathcal{B})}\Big\},\label{equ46}
\end{align}
then, $\mathbf{x}_k$ and ${\bm\lambda}_k$ generated by Algorithm \ref{alg1} linearly converge to $\mathbf{x}^*$ and ${\bm\lambda}^*$, respectively. Specifically,
\begin{align}\label{e47}
E(k+1)\leq aE(k),
\end{align}
where $a:=\max\{\rho(M_{\alpha}),1-\alpha\beta\lambda_{\min}(\Pi\Pi^{\top}),1-\underline{\sigma}^2(\mathcal{B})\gamma\}$ and $a\in(0,1)$.
\end{theorem}

\emph{Proof:}
It should be noticed that
\begin{align}
    &\|\tilde{\bm\lambda}_k-\mathcal{B}^2\tilde{\bm\lambda}_k+\beta\Pi(x_{k+1}-x^*)\|^2\nonumber\\
    &=\|\tilde{\bm\lambda}_k\|^2+\|-\mathcal{B}^2\tilde{\bm\lambda}_k+\beta\Pi(x_{k+1}-x^*)\|^2\nonumber\\
    &~~~+2\tilde{\bm\lambda}_k^{\top}(-\mathcal{B}^2\tilde{\bm\lambda}_k+\beta\Pi(x_{k+1}-x^*))\nonumber\\
    &\leq\|\tilde{\bm\lambda}_k\|^2+2\|\mathcal{B}^2\tilde{\bm\lambda}_k\|^2+2\beta^2\|\Pi(x_{k+1}-x^*)\|^2\nonumber\\
    &~~~-2\tilde{\bm\lambda}_k^{\top}\mathcal{B}^2\tilde{\bm\lambda}_k+2\beta\tilde{\bm\lambda}_k^{\top}\Pi(x_{k+1}-x^*).\label{equ47}
\end{align}
Also, by iteration (\ref{equ32}) and the property pf projection operator, it is seen that
\begin{align}
    \|\tilde{\bm\lambda}_{k+1}\|
    &=\|P_{\mathbb{R}^{Nm}_+}[\mathbf{v}_{k+1}]-P_{\mathbb{R}^{Nm}_+}[\mathbf{v}^*]\|\nonumber\\
    &\leq\|\tilde{\mathbf{v}}_{k+1}\|.\label{equ48}
\end{align}
Then it can be derived from Lemmas \ref{lemma2} and \ref{lemma3} along with (\ref{equ47}) and (\ref{equ48}) that
\begin{align}
    &\|\tilde{\mathbf{x}}_{k+1}\|^2+\alpha\beta^{-1}\|\tilde{\bm\lambda}_{k+1}\|^2_{I_{Nm}-\gamma\mathcal{B}^2}+\alpha\beta^{-1}\gamma^{-1}\|\tilde{\mathbf{y}}_{k+1}\|^2\nonumber\\
    &\leq\|\tilde{\mathbf{x}}_{k+1}\|^2+\alpha\beta^{-1}\|\tilde{\mathbf{v}}_{k+1}\|^2_{I_{Nm}-\gamma\mathcal{B}^2}+\alpha\beta^{-1}\gamma^{-1}\|\tilde{\mathbf{y}}_{k+1}\|^2\nonumber\\
    &\leq\rho(M_{\alpha})\|\tilde{\mathbf{x}}_k\|^2-\frac{\mu}{N}\alpha\|\mathcal{W}_{\infty}\mathbf{x}_k-\mathbf{x}^*\|^2
   -\sigma{L}\alpha\|\mathbf{x}_{\bot,k}\|^2\nonumber\\
   &~~~-\alpha^2\|\Pi^{\top}\tilde{\bm\lambda}_k\|^2+\alpha\beta^{-1}\|\tilde{\bm\lambda}_k\|^2+2\alpha\beta^{-1}\|\mathcal{B}^2\tilde{\bm\lambda}_k\|^2\nonumber\\
    &~~~-2\alpha\beta^{-1}\|\mathcal{B}\tilde{\bm\lambda}_k\|^2-\alpha\beta^{-1}\|\mathcal{B}\tilde{\mathbf{y}}_k\|^2+\alpha\beta^{-1}\gamma^{-1}\|\tilde{\mathbf{y}}_k\|^2\nonumber\\
    &~~~+2\alpha\beta\|\Pi\mathcal{R}(\mathbf{x}_{k+1}-\mathbf{x}^*)\|^2,
\end{align}
which implies that (\ref{equ45}) holds.

Moreover, for the first two terms on the right-hand side of (\ref{equ45}), one has
\begin{align}
    &\rho(M_{\alpha})\|\tilde{\mathbf{x}}_k\|^2-\min\{\mu/N,\sigma L\}\alpha\|\tilde{\mathbf{x}}_k\|^2\nonumber\\
    &=\rho(M_{\alpha})\|\tilde{\mathbf{x}}_k\|^2_{I_{Nn}-2\alpha\beta\mathcal{R}^{\top}\Pi^{\top}\Pi\mathcal{R}}\nonumber\\
    &~~~+\rho(M_{\alpha})\|\tilde{\mathbf{x}}_k\|^2_{2\alpha\beta\mathcal{R}^{\top}\Pi^{\top}\Pi\mathcal{R}}-\min\{\mu/N,\sigma L\}\alpha\|\tilde{\mathbf{x}}_k\|^2\nonumber\\
    &\leq\rho(M_{\alpha})\|\tilde{\mathbf{x}}_k\|^2_{I_{Nn}-2\alpha\beta\mathcal{R}^{\top}\Pi^{\top}\Pi\mathcal{R}}\nonumber\\
    &~~~+[2\alpha\beta\rho(M_{\alpha})\lambda_{\max}(\Pi^{\top}\Pi)-\min\{\mu/N,\sigma L\}\alpha]\|\tilde{\mathbf{x}}_k\|^2\nonumber\\
    &\leq\rho(M_{\alpha})\|\tilde{\mathbf{x}}_k\|^2_{I_{Nn}-2\alpha\beta\mathcal{R}^{\top}\Pi^{\top}\Pi\mathcal{R}},\label{equ51}
\end{align}
where the second inequality is obtained based on the selection of $\alpha$ and $\beta$ in (\ref{equ46}).

For the four terms about $\tilde{\bm\lambda}_k$ on the right-hand side of (\ref{equ45}), it can be derived that
\begin{align}
    &-\alpha^2\|\Pi^{\top}\tilde{\bm\lambda}_k\|^2+\alpha\beta^{-1}\|\tilde{\bm\lambda}_k\|^2+2\alpha\beta^{-1}\|\mathcal{B}^2\tilde{\bm\lambda}_k\|^2\nonumber\\
    &~-2\alpha\beta^{-1}\|\mathcal{B}\tilde{\bm\lambda}_k\|^2\nonumber\\
    &\leq\alpha\beta^{-1}(1-\alpha\beta\lambda_{\min}(\Pi\Pi^{\top}))\|\tilde{\bm\lambda}_k\|^2_{I_{Nm}-\gamma\mathcal{B}^2}\nonumber\\
    &~~~+\alpha\beta^{-1}\gamma(1-\alpha\beta\lambda_{\min}(\Pi\Pi^{\top}))\|\mathcal{B}\tilde{\bm\lambda}_k\|^2\nonumber\\
    &~~~+2\alpha\beta^{-1}(\lambda_{\max}^2(\mathcal{B})-1)\|\mathcal{B}\tilde{\bm\lambda}_k\|^2\nonumber\\
    &\leq\alpha\beta^{-1}(1-\alpha\beta\lambda_{\min}(\Pi\Pi^{\top}))\|\tilde{\bm\lambda}_k\|^2_{I_{Nm}-\gamma\mathcal{B}^2},\label{equ52}
\end{align}
where the second inequality depends on the selection of $\gamma$ in (\ref{equ46}).

Then, for the terms on $\tilde{\mathbf{y}}_k$ on the right-hand side of (\ref{equ45}), it holds that
\begin{align}
    &-\alpha\beta^{-1}\|\mathcal{B}\tilde{\mathbf{y}}_k\|^2+\alpha\beta^{-1}\gamma^{-1}\|\tilde{\mathbf{y}}_k\|^2\nonumber\\
    &\leq\alpha\beta^{-1}\gamma^{-1}(\|\tilde{\mathbf{y}}_k\|^2-\gamma\|\mathcal{B}\tilde{\mathbf{y}}_k\|^2)\nonumber\\
    &\leq\alpha\beta^{-1}\gamma^{-1}(1-\gamma\underline{\sigma}^2(\mathcal{B}))\|\tilde{\mathbf{y}}_k\|^2,\label{equ53}
\end{align}
where the last inequality is by $\|\mathcal{B}\tilde{\mathbf{y}}_k\|^2\geq\underline{\sigma}^2(\mathcal{B})\|\tilde{\mathbf{y}}_k\|^2$ under Assumption \ref{assumption3} since $\tilde{\mathbf{y}}_k$ is in the range space of $\mathcal{B}$ \cite{ren2008distributed}.

Combining (\ref{equ51})--(\ref{equ53}), one has (\ref{e47}) holds. Hence, the prove is completed.
\hfill$\blacksquare$

Theorem \ref{theorem1} presents the linear last-iterate convergence result on the designed distributed GNE seeking algorithm in Algorithm \ref{alg1} and also provides the explicit convergence rate by appropriately selecting stepsizes $\alpha,\beta$ and $\gamma$. Upper bounds of $\beta$ and $\gamma$ are provided in Theorem \ref{theorem1}, however, how to find a feasible stepsize $\alpha$ is still not clear.
In what follows, a corollary on $\alpha$ is derived to show the range of feasible $\alpha$ ensuring $\rho(M_{\alpha})<1$.
\begin{corollary}\label{corollary1}
If $\alpha>0$ satisfies
\begin{align}
     \alpha<\min\Big\{\frac{1-\sigma^2}{9\sigma{L}},\frac{\sqrt{1-\sigma^2}}{\sqrt{3}L},\frac{\mu(1-\sigma)}{6NL^2}\Big\},
\end{align}
then $\rho(M_{\alpha})<1$.
\end{corollary}

\emph{Proof:}
Note that $M_{\alpha}$ is a symmetric matrix, then $\rho(M_{\alpha})<1$ is equivalent to that $M_{\alpha}+I_2$ and $I_2-M_{\alpha}$ are positive definite matrices. By the Sylvester's criterion, $M_{\alpha}+I_2$ is positive definite if and only if the (1,1)th element of $M_{\alpha}+I_2$ is positive and the determinant of $M_{\alpha}+I_2$ is positive, that is,
\begin{align}
    &2-\frac{\mu}{N}\alpha+L^2\alpha^2>0,\label{equ55}\\
    &(2-\frac{\mu}{N}\alpha+L^2\alpha^2)(1+\sigma^2+3\sigma L\alpha+L^2\alpha^2)\nonumber\\
    &-(\sigma+1)^2L^2\alpha^2>0.\label{equ56}
\end{align}
It can be easily seen that (\ref{equ55}) holds if $2-\frac{\mu}{N}\alpha>0$, i.e.,
\begin{align}\label{equ57}
    \alpha<\frac{2N}{\mu}.
\end{align}
By a simple computation, (\ref{equ56}) can be rewritten as
\begin{align}
    &2(1+\sigma^2)+6\sigma L\alpha+2(1-\sigma)L^2\alpha^2-\frac{\mu(1+\sigma^2)}{N}\alpha\nonumber\\
    &-\frac{3\mu\sigma L}{N}\alpha^2-\frac{\mu L^2}{N}\alpha^3+3\sigma L^3\alpha^3+L^4\alpha^4>0,
\end{align}
which can be implied by
\begin{align*}
    &2(1+\sigma^2)-\frac{\mu(1+\sigma^2)}{N}\alpha>0,\\
    &6\sigma L\alpha-\frac{3\mu\sigma{L}}{N}\alpha^2>0,\\
    &2(1-\sigma)L^2\alpha^2-\frac{\mu L^2}{N}\alpha^3>0,
\end{align*}
that is,
\begin{align}\label{equ59}
    \alpha<\min\Big\{\frac{2N}{\mu},\frac{2N(1-\sigma)}{\mu}\Big\}.
\end{align}
Combining (\ref{equ57}) and (\ref{equ59}) yields that $M_{\alpha}+I_2$ is positive definite if
\begin{align}
    \alpha<\frac{2N(1-\sigma)}{\mu}.
\end{align}
By the Sylvester's criterion again and similar to the above process, one has that $I_2-M_{\alpha}$ is positive definite if $\alpha$ satisfies
\begin{align}
    \alpha<\min\Big\{\frac{\mu}{NL^2},\frac{1-\sigma^2}{9\sigma{L}},\frac{\sqrt{1-\sigma^2}}{\sqrt{3}L},\frac{\mu(1-\sigma)}{6NL^2}\Big\}.
\end{align}
It should be noticed that $\frac{2N(1-\sigma)}{\mu}>\frac{\mu(1-\sigma)}{6NL^2}$ since $\mu\leq L$ and $\frac{\mu}{NL^2}>\frac{\mu(1-\sigma)}{6NL^2}$.
Thus, the proof is completed. \hfill$\blacksquare$

\begin{remark}
From Theorem \ref{theorem1} and Corollary \ref{corollary1}, it can be observed that the upper bounds of stepsizes $\alpha,\beta$ and $\gamma$ depend on the number of the players, the communication structure, and the affine inequality, as well as the Lipschitz and monotonicity constants of the pseudo-gradient function. In fact, from the proofs, one can find that these bounds on the stepsizes are not tight and larger stepsizes may be selected to ensure better convergence results.
\end{remark}

\section{Simulations}   \label{section4}
\begin{figure}[!ht]
\centering
\includegraphics[width=2.8in]{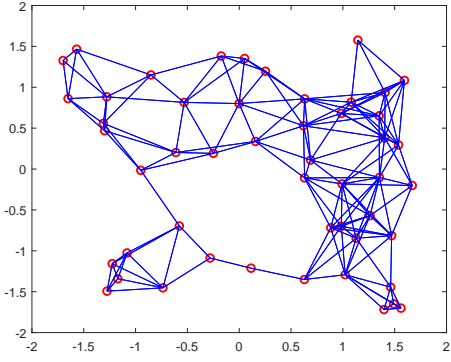}
\caption{Random communication graph with 50 nodes.}\label{figure0}
\end{figure}

\begin{figure}[!ht]
\centering
\includegraphics[width=2.5in]{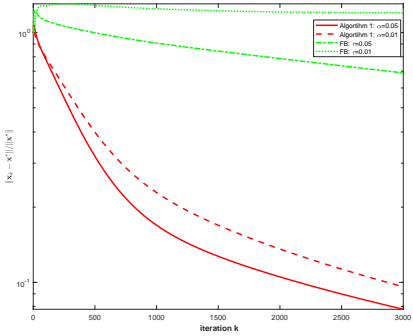}
\caption{Distance from the variational GNE for Algorithm \ref{alg1} and the FB algorithm in \cite{pavel2020distributed} under different stepsizes.}\label{figure1}
\end{figure}

In this section, consider a Nash-Cournot game \cite{pavel2020distributed}, where there are $N$ firms producing a commodity that is sold to $m$ markets. Each firm $i\in[N]$ participates in $n_i~(\leq m)$ of the $m$ markets and determines its production quantities $x_i\in\mathbb{R}^{n_i}$ to be delivered to the $n_i$ markets. Since each market has a maximal capacity, which results in a coupled affine inequality $Ax\leq b$, where $A=[A_1,\ldots,A_N]$ with $A\in\mathbb{R}^{m\times n_i}$, $x=col(x_1,\ldots,x_N)$, $b=col(b_1,\ldots,b_m)$ with $b_i>0$ being the capacity of the market $i$. The aim of each firm is to minimize its cost function $f_i(x_i,x_{-i})=c_i(x_i)-p(Ax)^{\top}A_ix_i$, where $c_i(x_i)=x_i^{\top}Q_ix_i+q_i^{\top}x_i$ is the production cost of firm $i$ with $Q_i\in\mathbb{R}^{n_i\times n_i}$ being a positive definite matrix, $q_i\in\mathbb{R}^{n_i}$ and $p:\mathbb{R}^m\to\mathbb{R}^m$ being a price vector function associating with the markets. Specifically, the price for market $i$ is the $i$th element of $p(Ax)$, i.e., $[p(Ax)]_i=L_i-w_i[Ax]_i$, where $L_i>0$ and $w_i>0$.
Set $N=50$, $m=n_i=5$, and $A_i=I_m$, $i\in[N]$. Choose $Q_i$ to be a diagonal matrix. For each $i\in[N]$, randomly select the diagonals of $Q_i$, $b_i$, $L_i$ and $w_i$ from $[1,8]$, $[1,2]$, $[10,20]$, $[1,3]$ and $[5,10]$ with uniform distributions, respectively. The communication graph is randomly created as shown in Fig. \ref{figure0}.
This setup satisfies all our theoretical assumptions, and set the stepsizes $\beta=0.1$ and $\gamma=0.1$. The evolutions of $\|{\bf x}_{k}-{\bf x}^*\|$ generated by Algorithm \ref{alg1} and the proposed forward-backward (FB) algorithm in \cite{pavel2020distributed} are provided in Fig. \ref{figure1} under different stepsizes $\alpha$ and $\tau$, respectively, where two algorithms are performed with the same initial conditions and implemented by Matlab R2020a running on a laptop equipped with Intel(R) Core(TM) i7-1065G7 CPU @ 1.30GHz. It can be seen from Fig. \ref{figure1} that our algorithm performs a faster convergence, and larger feasible stepsizes also lead to the faster convergence.

\section{Conclusion}\label{section5}

In this paper, the GNE seeking problem for continuous games with coupled affine inequality constraints was solved by designing a novel primal-dual algorithm. The linear last-iterate convergence of the designed algorithm was also rigorously analyzed and the bounds of feasible stepsizes were provided. Future work of interest may be on the design of linearly convergent algorithms for continuous games with compact strategy set constraints and coupled nonlinear inequality constraints.

\section*{Appendix}

\subsection{Proof of Lemma \ref{lemma1}}

Assume that the variational GNE $x^*$ and the global dual variable $\lambda^*$ are given, which satisfy (\ref{equ4}) and (\ref{equ5}) with $\lambda_1^*=\lambda_2^*=\cdots=\lambda_N^*=\lambda^*$. Let ${\bf x}^*={\bf1}_N\otimes{x}^*$ and ${\bm\lambda}^*={\bf1}_N\otimes\lambda^*$, then it holds that $\mathbf{x}^*=(W\otimes I_n)\mathbf{x}^*$ and $\mathcal{B}^2{\bm\lambda}^*={\bf0}_{Nm}$ under Assumption \ref{assumption3}. In view of $\mathbf{F}(\mathbf{x}^*)=F(x^*)$ by the definitions in (\ref{equ2}) and (\ref{equ17}), it can be obtained from (\ref{equ4}) that
\begin{align*}
    \mathbf{F}(\mathbf{x}^*)+\Pi^{\top}{\bm\lambda}^*={\bf0}_{n}.
\end{align*}
Thus, $\mathbf{x}^*$ and ${\bm\lambda}^*$ satisfy (\ref{equ19}). Define
\begin{align}
    \mathbf{v}^*=\mathbf{1}_N\otimes{v^*}=\mathbf{1}_N\otimes\Big(\lambda^*+\frac{\beta}{N}(Ax^*-b)\Big),
\end{align}
then $\mathbf{v}^*$ satisfies (\ref{equ21}). Since (\ref{equ22}) is equivalent to $\lambda^*=P_{\mathbb{R}^m_+}[v^*]$, that is,
\begin{align}
    (v^*-\lambda^*)^{\top}(v-\lambda^*)\leq0,~\forall v\in\mathbb{R}^m_+,
\end{align}
i.e.,
\begin{align}
    \frac{\beta}{N}(Ax^*-b)^{\top}(v-\lambda^*)\leq0,~\forall v\in\mathbb{R}^m_+,
\end{align}
one has that (\ref{equ22}) holds by (\ref{equ5}). Consequently, it remains to prove that there exists $\mathbf{y}^*$ satisfying (\ref{equ20}). As $(\mathbf{1}^{\top}_N\otimes I_m)(\mathbf{v}^*-{\bm\lambda}^*-\beta(\Pi x^*-\mathbf{b}))={\bf0}$, it can be seen that
$\mathcal{B}\mathbf{y}^*$ is in the null space of $\mathbf{1}^{\top}_N\otimes I_m$ and thus is in the range space of $\mathcal{B}^2$. Subsequently, there exists $\mathbf{y}^*$ in the range space of $\mathcal{B}$ satisfying (\ref{equ20}).

On the other hand, suppose that $({\bf x}^*,\mathbf{v}^*,\mathbf{y}^*,{\bm\lambda}^*)$ is a fixed point of iteration (\ref{equ7}), then it can be derived from (\ref{equ21}) and $\mathcal{B}^2\bm{\lambda}^*={\bf0}_{Nm}$ that $\mathbf{v}^*=\mathbf{1}_N\otimes v^*$ for some $v^*\in\mathbb{R}^m$ and ${\bm\lambda}^*=\mathbf{1}_N\otimes{\lambda}^*$ for some $\lambda^*\in\mathbb{R}^m$. Multiplying ${\bf1}_N^{\top}\otimes{I_n}$ on both sides of (\ref{equ19}) yields
\begin{align}\label{equ26}
   \mathbf{F}(\mathbf{x}^*)+\Pi^{\top}{\bm\lambda}^*=\mathbf{0}_n,
\end{align}
where $\mathbf{1}^{\top}_NW=\mathbf{1}^{\top}_N$ is applied. Therefore, substituting (\ref{equ26}) into (\ref{equ19}) yields $\mathbf{x}^*=(W\otimes I_n)\mathbf{x}^*$. Then, it is obtained from Assumption \ref{assumption3} that $\mathbf{x}^*=\mathbf{1}_N\otimes{x}^*$ for some $x^*\in\mathbb{R}^n$. Subsequently, by (\ref{equ19}), (\ref{equ20}) and (\ref{equ22}), it can be concluded that $x^*$ is the variational GNE of game (\ref{equ1}) and $\lambda^*$ is the optimal dual variable if selecting $\mathbf{y}^*$ from the range space of $\mathcal{B}$.
\hfill$\blacksquare$

\subsection{Proof of Lemma \ref{lemma2}}

Note that
\begin{align}
    &\|\tilde{\mathbf{x}}_{k+1}\|^2-\|\mathcal{W}\mathbf{x}_k-\mathbf{x}^*\|^2\nonumber\\
    &=-\|\mathbf{x}_{k+1}-\mathcal{W}\mathbf{x}_k\|^2+2\tilde{\mathbf{x}}_{k+1}^{\top}(\mathbf{x}_{k+1}-\mathcal{W}\mathbf{x}_k),\label{equ37}
\end{align}
where $a^2-(a-b)^2=-b^2+2ab$ is applied. Then, combining (\ref{equ35}), (\ref{equ37}), iteration (\ref{equ29}) and $\mathbf{x}_{k+1}-\mathcal{W}\mathbf{x}_k=\tilde{\mathbf{x}}_{k+1}-\mathcal{W}\tilde{\mathbf{x}}_k$, it can be derived that
\begin{align}
    \|\tilde{\mathbf{x}}_{k+1}\|^2&=\|\mathcal{W}_{\infty}\mathbf{x}_k-\mathbf{x}^*\|^2+\|\mathcal{W}\mathbf{x}_{\bot,k}\|^2-\|\mathbf{x}_{k+1}-\mathcal{W}\mathbf{x}_k\|^2\nonumber\\
    &~~~-2\alpha\tilde{\mathbf{x}}_{k+1}^{\top}\mathcal{R}^{\top}(\mathbf{F}(\mathbf{x}_k)-\mathbf{F}(\mathbf{x}^*))\nonumber\\
    &~~~-2\alpha\tilde{\mathbf{x}}_{k+1}^{\top}\mathcal{R}^{\top}\Pi^{\top}\tilde{\bm\lambda}_k.\label{equ38}
\end{align}
Based on Assumption \ref{assumption1}, the following inequality holds:
\begin{align*}
    \|\mathbf{F}(\mathbf{x})-\mathbf{F}(\mathbf{y})\|^2
    &=\sum_{i=1}^N\|\nabla_if_i(\mathbf{x}_i)-\nabla_if_i(\mathbf{y}_i)\|^2\\
    &\leq L^2\sum_{i=1}^N\|\mathbf{x}_i-\mathbf{y}_i\|^2\\
    &=L^2\|\mathbf{x}-\mathbf{y}\|^2
\end{align*} for any $\mathbf{x}=col(\mathbf{x}_1,\ldots,\mathbf{x}_N)$ and $\mathbf{y}=col(\mathbf{y}_1,\ldots,\mathbf{y}_N)$ with $\mathbf{x}_i,\mathbf{y}_i\in\mathbb{R}^n$. Hence,
\begin{align}
    &-2\alpha\tilde{\mathbf{x}}_{k+1}^{\top}\mathcal{R}^{\top}(\mathbf{F}(\mathbf{x}_k)-\mathbf{F}(\mathbf{x}^*))\nonumber\\
    &=-2\alpha(\mathcal{W}_{\infty}{\mathbf{x}}_{k}-\mathbf{x}^*)^{\top}\mathcal{R}^{\top}(\mathbf{F}(\mathcal{W}_{\infty}\mathbf{x}_k)-\mathbf{F}(\mathbf{x}^*))\nonumber\\
    &~~~-2\alpha(\mathcal{W}_{\infty}{\mathbf{x}}_{k}-\mathbf{x}^*)^{\top}\mathcal{R}^{\top}(\mathbf{F}(\mathbf{x}_k)-\mathbf{F}(\mathcal{W}_{\infty}\mathbf{x}_k))\nonumber\\
    &~~~-2\alpha(\mathcal{W}{\mathbf{x}}_{k}-\mathcal{W}_{\infty}\mathbf{x}_k)^{\top}\mathcal{R}^{\top}(\mathbf{F}(\mathbf{x}_k)-\mathbf{F}(\mathbf{x}^*))\nonumber\\
    &~~~-2\alpha({\mathbf{x}}_{k+1}-\mathcal{W}\mathbf{x}_k)^{\top}\mathcal{R}^{\top}(\mathbf{F}(\mathbf{x}_k)-\mathbf{F}(\mathbf{x}^*))\nonumber\\
    &\leq-2\alpha\mu\|({\bf1}_N^{\top}\otimes I_n)\mathbf{x}_k/N-x^*\|^2\nonumber\\
    &~~~+2\alpha\|\mathcal{W}_{\infty}{\mathbf{x}}_{k}-\mathbf{x}^*\|\cdot L\|\mathbf{x}_k-\mathcal{W}_{\infty}\mathbf{x}_k\|\nonumber\\
    &~~~+2\alpha\|\mathcal{W}{\mathbf{x}}_{k}-\mathcal{W}_{\infty}\mathbf{x}_k\|\cdot L\|\mathbf{x}_k-\mathbf{x}^*\|\nonumber\\
    &~~~+\|\mathbf{x}_{k+1}-\mathcal{W}\mathbf{x}_k\|^2+\alpha^2\|\mathcal{R}^{\top}(\mathbf{F}(\mathbf{x}_k)-\mathbf{F}(\mathbf{x}^*))\|^2\nonumber\\
    &~~~-\|\mathbf{x}_{k+1}-\mathcal{W}\mathbf{x}_k+\alpha\mathcal{R}^{\top}(\mathbf{F}(\mathbf{x}_k)-\mathbf{F}(\mathbf{x}^*))\|^2\nonumber\\
    &\leq-\frac{2\mu}{N}\alpha\|\mathcal{W}_{\infty}\mathbf{x}_k-\mathbf{x}^*\|^2+2L\alpha\|\mathcal{W}_{\infty}{\mathbf{x}}_{k}-\mathbf{x}^*\|\cdot \|\mathbf{x}_{\bot,k}\|\nonumber\\
    &~~~+2L\alpha\|\mathcal{W}{\mathbf{x}}_{k}-\mathcal{W}_{\infty}\mathbf{x}_k\|\cdot \|\mathbf{x}_k-\mathbf{x}^*\|\nonumber\\
    &~~~+\|\mathbf{x}_{k+1}-\mathcal{W}\mathbf{x}_k\|^2+L^2\alpha^2\|\mathbf{x}_k-\mathbf{x}^*\|^2\nonumber\\
    &~~~-\alpha^2\|\mathcal{R}^{\top}\Pi^{\top}\tilde{\bm\lambda}_k\|^2,
\end{align}
where the first inequality is obtained based on Assumption \ref{assumption2}, $\|\mathcal{R}\|\leq 1$ and $-2ab=a^2+b^2-(a+b)^2$, and the second inequality is derived by (\ref{equ18}) and $\mathbf{F}(\mathbf{x}^*)+\Pi^{\top}{\bm\lambda}^*={\bf0}_{n}$.
As a consequence, one can rewrite (\ref{equ38}) as
\begin{align}
    &\|\tilde{\mathbf{x}}_{k+1}\|^2\nonumber\\
    &\leq\|\mathcal{W}_{\infty}\mathbf{x}_k-\mathbf{x}^*\|^2+\|\mathcal{W}\mathbf{x}_{\bot,k}\|^2-\frac{2\mu}{N}\alpha\|\mathcal{W}_{\infty}\mathbf{x}_k-\mathbf{x}^*\|^2\nonumber\\
    &~~~+2L\alpha\|\mathcal{W}_{\infty}\mathbf{x}_k-\mathbf{x}^*\|\cdot\|\mathbf{x}_{\bot,k}\|\nonumber\\
    &~~~+2L\alpha\|\mathcal{W}\mathbf{x}_k-\mathcal{W}_{\infty}\mathbf{x}_k\|\cdot\|\mathbf{x}_k-\mathbf{x}^*\|+L^2\alpha^2\|\mathbf{x}_k-\mathbf{x}^*\|^2\nonumber\\
    &~~~-\alpha^2\|\mathcal{R}^{\top}\Pi^{\top}\tilde{\bm\lambda}_k\|^2-2\alpha\tilde{\mathbf{x}}_{k+1}^{\top}\mathcal{R}^{\top}\Pi^{\top}\tilde{\bm\lambda}_k\nonumber\\
    &\leq\|\mathcal{W}_{\infty}\mathbf{x}_k-\mathbf{x}^*\|^2+\sigma^2\|\mathbf{x}_{\bot,k}\|^2-\frac{2\mu}{N}\alpha\|\mathcal{W}_{\infty}\mathbf{x}_k-\mathbf{x}^*\|^2\nonumber\\
    &~~~+2L\alpha\|\mathcal{W}_{\infty}\mathbf{x}_k-\mathbf{x}^*\|\cdot\|\mathbf{x}_{\bot,k}\|\nonumber\\
    &~~~+2L\alpha\sigma\|\mathbf{x}_{\bot,k}\|\cdot(\|\mathcal{W}_{\infty}\mathbf{x}_k-\mathbf{x}^*\|+\|\mathbf{x}_{\bot,k}\|)\nonumber\\
    &~~~+L^2\alpha^2(\|\mathcal{W}_{\infty}\mathbf{x}_k-\mathbf{x}^*\|^2+\|\mathbf{x}_{\bot,k}\|^2)\nonumber\\
    &~~~-\alpha^2\|\mathcal{R}^{\top}\Pi^{\top}\tilde{\bm\lambda}_k\|^2-2\alpha(x_{k+1}-x^*)^{\top}\Pi^{\top}\tilde{\bm\lambda}_k\nonumber\\
    &\leq(\|\mathcal{W}_{\infty}\mathbf{x}_k-\mathbf{x}^*\|,\|\mathbf{x}_{\bot,k}\|){M_{\alpha}}(\|\mathcal{W}_{\infty}\mathbf{x}_k-\mathbf{x}^*\|,\|\mathbf{x}_{\bot,k}\|)^{\top}\nonumber\\
    &~~~-\frac{\mu}{N}\alpha\|\mathcal{W}_{\infty}\mathbf{x}_k-\mathbf{x}^*\|^2-\sigma{L}\alpha\|\mathbf{x}_{\bot,k}\|^2\nonumber\\
    &~~~-\alpha^2\|\mathcal{R}^{\top}\Pi^{\top}\tilde{\bm\lambda}_k\|^2-2\alpha(x_{k+1}-x^*)^{\top}\Pi^{\top}\tilde{\bm\lambda}_k,
\end{align}
where the second inequality is derived based on $\|\mathcal{W}\mathbf{x}_k-\mathcal{W}_{\infty}\mathbf{x}_k\|=\|\mathcal{W}{\bf x}_{\bot,k}\|\leq\sigma\|{\bf x}_{\bot,k}\|$ and $ \|\mathbf{x}_k-\mathbf{x}^*\|^2=\|\mathcal{W}_{\infty}\mathbf{x}_k-\mathbf{x}^*\|^2+\|\mathbf{x}_{\bot,k}\|^2$.
Thus, the lemma is proved. \hfill$\blacksquare$

\subsection{Proof of Lemma \ref{lemma3}}

From (\ref{equ30}) and (\ref{equ31}), one has
\begin{align}
    \|\tilde{\mathbf{v}}_{k+1}\|^2
    &=\|\tilde{\bm\lambda}_k-\mathcal{B}^2\tilde{\bm\lambda}_k+\beta\Pi(x_{k+1}-x^*)+\mathcal{B}\tilde{\mathbf{y}}_k\|^2\nonumber\\
    &=\|\tilde{\bm\lambda}_k-\mathcal{B}^2\tilde{\bm\lambda}_k+\beta\Pi(x_{k+1}-x^*)\|^2+\|\mathcal{B}\tilde{\mathbf{y}}_k\|^2\nonumber\\
    &~~~+2(\tilde{\bm\lambda}_k-\mathcal{B}^2\tilde{\bm\lambda}_k+\beta\Pi(x_{k+1}-x^*))^{\top}\mathcal{B}\tilde{\mathbf{y}}_k,
\end{align}
and
\begin{align}
    \|\tilde{\mathbf{y}}_{k+1}\|^2
    &=\|\tilde{\mathbf{y}}_k-\gamma\mathcal{B}\tilde{\mathbf{v}}_{k+1}\|^2\nonumber\\
    &=\|\tilde{\mathbf{y}}_k\|^2+\gamma^2\|\mathcal{B}\tilde{\mathbf{v}}_{k+1}\|^2-2\gamma\tilde{\mathbf{y}}_k^{\top}\mathcal{B}\tilde{\mathbf{v}}_{k+1}\nonumber\\
    &=\|\tilde{\mathbf{y}}_k\|^2+\gamma^2\|\mathcal{B}\tilde{\mathbf{v}}_{k+1}\|^2-2\gamma\tilde{\mathbf{y}}_k^{\top}\mathcal{B}^2\tilde{\mathbf{y}}_k\nonumber\\
    &~~~-2\gamma\tilde{\mathbf{y}}_k^{\top}\mathcal{B}(\tilde{\bm\lambda}_k-\mathcal{B}^2\tilde{\bm\lambda}_k+\beta\Pi(x_{k+1}-x^*)).
\end{align}
Combining the above two equations yields
\begin{align}
   &\|\tilde{\mathbf{v}}_{k+1}\|^2+\gamma^{-1}\|\tilde{\mathbf{y}}_{k+1}\|^2\nonumber\\
   &=\|\tilde{\bm\lambda}_k-\mathcal{B}^2\tilde{\bm\lambda}_k+\beta\Pi(x_{k+1}-x^*)\|^2-\|\mathcal{B}\tilde{\mathbf{y}}_k\|^2\nonumber\\
   &~~~+\gamma^{-1}\|\tilde{\mathbf{y}}_k\|^2+\gamma\|\mathcal{B}\tilde{\mathbf{v}}_{k+1}\|^2.
\end{align}
Then, a simple computation leads to (\ref{equ41}). \hfill$\blacksquare$

\bibliographystyle{IEEEtran}

\end{document}